# SOT-MRAM 300mm integration for low power and ultrafast embedded memories

K. Garello[1], F. Yasin[1], S. Couet[1], L. Souriau[1], J. Swerts[1], S. Rao[1], S. Van Beek[1], W. Kim[1], E. Liu[1], S. Kundu[1], D. Tsvetanova[1], N. Jossart[1], K. Croes[1], E. Grimaldi[2], M. Baumgartner[2], D. Crotti[1], A. Furnémont[1], P. Gambardella[2], G.S. Kar[1]

[1]imec, 3001 Leuven, Belgium ; [2]Department of Materials, ETHZ, 8093 Zürich, Switzerland; email: kevin.garello@imec.be

**Abstract** — We demonstrate for the first time full-scale integration of top-pinned perpendicular MTJ on 300 mm wafer using CMOS-compatible processes for spin-orbit torque (SOT)-MRAM architectures. We show that 62 nm devices with a W-based SOT underlayer have very large endurance (> $5\times10^{10}$), sub-ns switching time of 210 ps, and operate with power as low as 300 pJ.

**Introduction:** The introduction of non-volatility (NV) at the cache level in advance logic nodes is sought as it would lead to a large decrease of the power consumption of microprocessors. Among NV memory technologies, spin-transfer torque (STT) MRAM has gained a lot of attention due to its scalability, low power and high speed, as well as compatibility with scaled CMOS processes and voltages. Despite all these advantages, STT-MRAM cannot operate reliably at ns and sub-ns scales due to large incubation delays [1, 2], making it an unsuitable solution to tackle L1/2 SRAM cache replacement. In addition, the shared read/write path can impair the read reliability, while the write current can impose severe stress on the MTJ, leading to time dependent degradation of the memory cell. To mitigate these issues, spin-orbit torque (SOT)-MRAM has been recently proposed [2, 3]. SOT induces switching of the free layer (FL) of the MTJ by injecting an in-plane current in an adjacent SOT layer, typically with the assistance of a static in-plane magnetic field [2]. This enables a three terminal MTJ-based concept that isolates the read/write path (Fig. 1), significantly improving the device endurance and read stability. Moreover, due to SOT spin transfer geometry, incubation time is negligible which allows for reliable switching operation at sub-ns timescales [4, 5]. Here, we report the first successful integration of SOT-MTJ cells on 300 mm wafers using CMOS-compatible processes. We demonstrate low power sub-ns switching and pathways for further optimization. Finally, excellent endurance and absence of electro-migration effect of ultrathin SOT layers are shown.

**Integration flow**: We used a SOT dedicated mask set in the imec 300 mm fab. The main steps of the integration process are summarized in Fig. 2: a SOT-MTJ stack is deposited on smooth bottom electrodes (BE), which are fabricated using a tungsten (W) damascene process. The MTJ is top pinned and consist of SOT/CoFeB/MgO/CoFeB/SAF perpendicularly magnetized (PMA) stack, where the SOT layer is W-based. Specific stop etch conditions have been developed to leave the SOT layer intact while patterning the MTJ pillar without producing sidewall shorts across the MgO barrier (Fig. 2c,d). Subsequently, the SOT layer is etched to form the three terminal device and a dual damascene Cu top electrode (TE) was fabricated to complete the electrical connection (Fig. 2a).

**Stack development:** SOTs possess a damping-like term ($\tau_{DL}$) attributed to spin Hall and a field-like term ($\tau_{FL}$) attributed to interface interactions [2]. Recent work indicates that $\tau_{DL}$ triggers switching while $\tau_{FL}$ accelerates it [5]. Charge-to-spin conversion efficiency parameters $\theta_{DL}$ and $\theta_{FL}$ are commonly used to compare different SOT metals [2], and beta-phase W has the largest reported efficiency with $\theta_{DL}$ = -30% [6]. It is therfore a natural choice to develop our technology using $W_\beta$ seed as SOT layer. Since SOT are thickness dependent [2, 6], the first step optimizes the base of the MTJ stack that includes SOT and FL layers as a function of W thickness $t_W$: W($t_W$)/CoFeB(1nm)/MgO/Cap. In Fig. 3, we report the FL magnetization ($M_s$) and effective anisotropy field ($B_k$), as well as the stack resistivity ($\rho_{stack}$) and $\theta_{DL}$ dependence on $t_W$, finding optimum properties at ~$t_W$=4 nm, with $\theta_{DL}$ = -32% ($\theta_{FL}$ = -10%), in agreement with literature [6]. Above 4 nm, W turns into α-phase, which dramatically reduces $\theta_{DL}$ (-2%), stack resistivity (from 120 to 40 μΩ.cm), and leads to a perpendicular-to-in-plane magnetic anisotropy transition. We however show the possibility to extend the β-phase and the PMA window while maintaining large $\theta_{DL}$ by stack engineering W(II) (Fig. 3). Subsequently, we developed a top-pinned W-MTJ with optimized SAF to minimize offset field $B_{off}$ at small device dimensions (Fig. 4). Blanket TMR is 100% and resistance area (RA) is 10 Ω.μm².

**Device performances**: The TMR, coercivity field $B_c$, $B_k$ and the thermal stability Δ dependence on the MTJ electrical size (eCD) are resumed Fig. 5. It shows that our $W_\beta$-MTJ has already very good properties, close to state-of-art top-pinned stacks. SOT track resistance is typically ~330 Ω, equivalent to $\rho_{SOT}$~140 μΩ.cm, which confirms the excellent stop etch conditions of the MTJ. In the following, we discuss results obtained on devices of eCD = 62 nm, sitting on a 210 nm wide SOT track. We measure a TMR of 90%, $B_k$ > 250 mT, $B_c$ ~100 mT, $B_{off}$ ~10 mT and Δ~50. Secondly, using the *exactly same 62nm device*, we evaluate STT and SOT switching performances as a function of voltage pulse length and amplitude (Fig. 6). STT is measured at 0 external field while SOT switching requires less than 23 mT in-plane field. In STT case, $\tau_p$ is limited to 5ns to avoid breakdown and the critical switching voltage evaluated at 50% probability $V_{c,STT}^{50\%}$ is ~1V to switch both P and AP states, corresponding to an energy of ~ 470 fJ. In comparison, SOT demonstrates reliable switching down to 210 ps with $V_{c,SOT}^{50\%}$~1V, and 0.7 V at 500 ps. Despite large current densities (currently ~200 MA.cm⁻²), SOT switching is energy wise very favorable with a value as low as 350 fJ at 280 ps. This is due to the current injection geometry and to the low resistance of SOT track. Further reduction in energy at sub-ns can be obtained by current density reduction via design optimization and stack engineering. We finally show that the SOT track is extremely robust by performing endurance measurement on 50 nm device up to $5\times10^{10}$ events using 100 ns / 100 MA.cm⁻² pulses (Fig. 7). We continue with electro-migration test at 300°C in the presence of a dc current density of 40 MA.cm⁻² without signs of SOT and MTJ degradation. Finally, we resume in table 1 our W-STT and W-SOT device properties and compare them to a Ta-SOT MTJ [4].

**Conclusion**: We demonstrated for the first-time the possibility to fabricate state-of-the-art SOT-MRAM devices using CMOS-compatible, large scale integration processes. The sub-ns switching speed and the low power device performances show that SOT-MRAM has the capacity to tackle SRAM cache replacement. Further research is focused on reducing the current density and field-free switching via SOT layer optimization, stack engineering and device design.

**Acknowledgment:** This work was performed as part of imec's industrial affiliation program on MRAM devices and partially supported by the Swiss National Science Foundation (grant number 200020-172775).

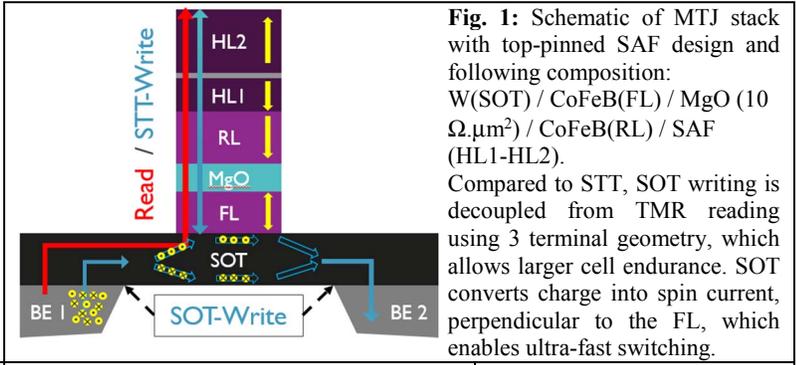

**Fig. 1:** Schematic of MTJ stack with top-pinned SAF design and following composition: W(SOT) / CoFeB(FL) / MgO (10 $\Omega.\mu m^2$) / CoFeB(RL) / SAF (HL1-HL2).
Compared to STT, SOT writing is decoupled from TMR reading using 3 terminal geometry, which allows larger cell endurance. SOT converts charge into spin current, perpendicular to the FL, which enables ultra-fast switching.

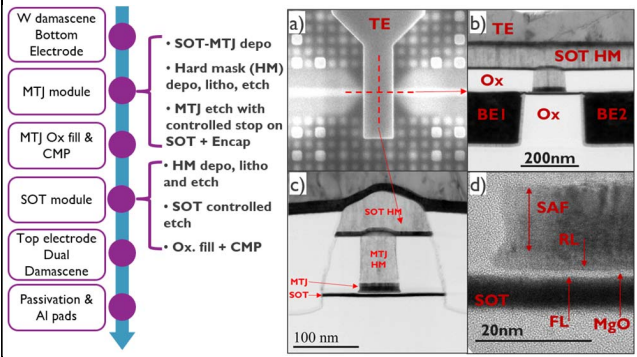

**Fig. 2:** Left chart resumes the main integration modules of SOT-MTJ technology. a) SEM top view of a SOT device, b/c) lateral/vertical TEM cross section views of the pillar showing a MTJ with straight profile sitting on a continuous SOT layer contacting 2 smooth BE, d) High resolution TEM on pillar edge of (c) highlighting sidewalls clean of shorts across the MgO barrier and reflecting the excellent stop MTJ etch condition stabilized.

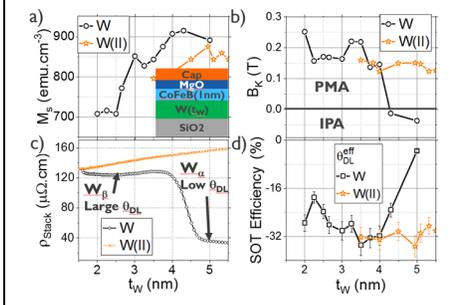

**Fig. 3:** SOT/FL optimization as a function of W thickness tw: a) $M_s$ of CoFeB free layer, b) Effective anisotropy $B_k$, c) Stack resistivity, d) SOT efficiency. For W (black), we have transition from β to α phase at ~4nm with PMA loss and strong $\theta_{DL}$ decrease above 4nm due to $W_\alpha$ phase, whereas our W(II) process (orange) allows enlarging β-phase window while maintaining large $\theta_{DL}$ and PMA.

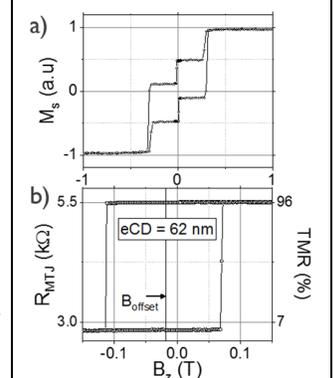

**Fig. 4:** a) Easy axis magnetization hysteresis loop of the unpatterned MTJ stack used in integrated devices, b) Free layer easy axis R-H hysterisis of a 62nm device.

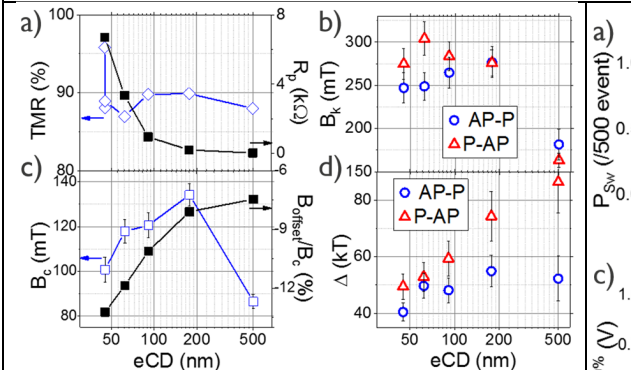

**Fig. 5:** MTJ properties vs. electrical size (eCD) for a) TMR and parallel state resistance, b) coercivity and $B_c/B_{off}$ ratio, c) anisotropy of P and AP states, d) thermal stability of P and AP states. It has been quantified using $B_c$ distribution probability method at room temperature.

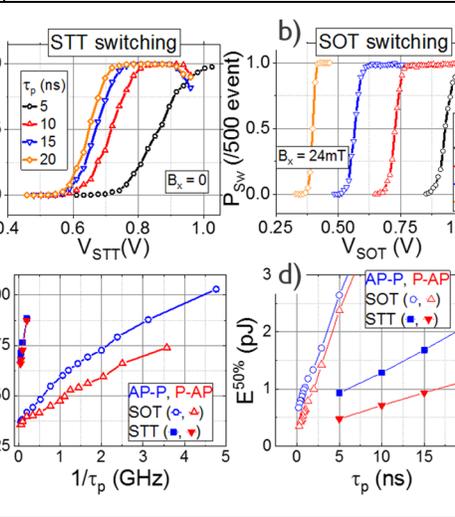

**Fig. 6:** STT and SOT switching operated separately on *exact same 62nm device*: a,b) examples of switching distribution as a function of pulse voltage $V_{STT,SOT}$ for various pulse length $\tau_p$ (P-AP transition), c) $V_c^{50\%}$ vs. inverse of pulse length showing switching up to 5 GHz. Remark that the offset field generates an imbalance of $V_{c,SOT}^{50\%}$ up to 15% between P-AP and AP-P transitions, d) Energy vs. time showing ultrafast and low power SOT-device performances.

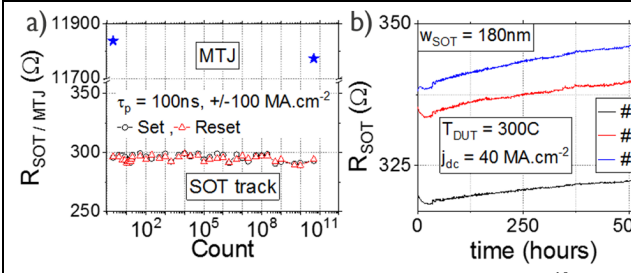

**Fig. 7:** SOT track endurance of a 50 nm MTJ. a) 5x $10^{10}$ cycles using as much energy as required for switching at 5 ns, b) Electro-migration test at 300°C/500 hours in the presence of dc current density of 40 MA.cm$^{-2}$ for 3 different samples. No device/TMR degradation is observed in both cases.

|  | W-STT | W-SOT | Ta-SOT[4] |
|---|---|---|---|
| eCD (nm) | 62 | 62 | 300 |
| TMR (%) / RA ($\Omega.\mu m^2$) | 90 / 12 | 90 / 12 | 55 / 600 |
| $H_c$ / $H_{offset}$ (mT) | 100 / 12 | 100 / 12 | ~ 50/5 |
| $B_k$ (T) | 0.29 | 0.29 | 0.5 |
| Δ | 46 | 46 |  |
| $\tau_p$ (ps) | 5000 | 210 | 400 |
| E @ 5 / 0.5 ns (pJ) | 0.47 / ? | 2.36 / 0.45 | ~4 / 8 |
| SOT Cycling |  | >5x$10^{10}$ |  |
| $\theta_{DL/FL}^{eff}$ (%) | Nan | -32 / -10 | -8 / -10 |
| $\rho_{SOT}$ ($\mu\Omega.cm$) | Nan | 140 | 160 |

**Table 1:** Scorecard of MTJs properties and writing performances comparing W-STT, W-SOT (same device) and Ta-SOT [4] top pinned based stack.